\documentclass[12pt]{article}
\usepackage{graphicx}
\usepackage[scale=0.8]{geometry}
\title{Wigner functions in quantum mechanics with a minimum length scale arising from generalized uncertainty principle}
\author{Prathamesh Yeole, Vipul Kumar, Kaushik Bhattacharya
\thanks{ E-mail:~ prathamy@iitk.ac.in, vipulku@iitk.ac.in, kaushikb@iitk.ac.in} 
\\ 
\normalsize
Department of Physics,\\
\normalsize
Indian Institute of Technology, Kanpur\\ 
\normalsize
Kanpur 208016, India}
\begin{document}

\maketitle

\begin{abstract}
In this paper we generalize the concept of Wigner function in the case
of quantum mechanics with a minimum length scale arising due to the
application of a generalized uncertainty principle (GUP). We present
the phase space formulation of such theories following GUP and show
that the Weyl transform and the Wigner function satisfy most of
their known properties in standard quantum mechanics. We utilize the
generalized Wigner function to calculate the phase space average of
the Hamiltonian of a quantum harmonic oscillator satisfying deformed
Heisenberg algebra. It is also shown that averages of certain quantum
mechanical operators in such theories may restrict the value of the
deformation parameter specifying the degree of deformation of
Heisenberg algebra. All the results presented are for pure states. The
results can be generalized for mixed states.
\end{abstract}

%%%%%%%%%%%%%%%%%%%%%%%%%%%%%%%%%%%%%%%%%%%%%%%%%%%%%%%%%%%%%%%%%%%%%%%%%%%%%%
\section{Introduction}

In general, quantum mechanics (QM) is not formulated in phase space,
standard QM formulation uses either the position space, or the
momentum space where the probability density functions are written as
$|\psi(x)|^{2} $ or $|\phi(p)|^{2}$ where $\psi(x)=\langle x|
\psi\rangle$ and $\phi(p)=\langle p| \psi\rangle$ are the solutions of
Schroedinger's equation, in one dimension, in the position space or
momentum space.  A phase space formulation of QM can be done with the
help of Wigner function, a function which depends both on $x$ and
$p$. Although, Wigner function does not exactly give a probability
distribution as it may have negative values, it does give a phase
space description of QM, which is why it is referred  as a 
quasi-distribution function \cite{case}.  In the present work we will
attempt to generalize the definition of the Wigner function in the
case where the standard Heisenberg uncertainty principle in QM has
been modified to a generalized uncertainty principle (GUP) as given in
\cite{kempf}. The modification of the uncertainty principle certainly
affects the phase space description of QM in an interesting manner.

In standard quantum mechanics there is no restriction on observable
length. We can see this by acknowledging that we can have $\Delta x
\rightarrow 0$ when $\Delta p \rightarrow \infty$ from Heisenberg's
uncertainty principle  $\Delta x\Delta p \geq \hbar/2$. But theories in quantum gravity suggest that
there must be a minimal observable length
\cite{maggore}. Theoretically we call it the minimal length
uncertainty $\Delta x_{0}$. 

In Ref.~\cite{kempf} the authors even
foresee that an effective theory of compound particles may be described
by QM with a minimal length scale. In the present paper we present the
phase space picture of QM with a minimal length scale following the
work done in Ref.~\cite{kempf}. In the above mentioned work the authors have presented the Hilbert space representation of QM with a minimal length scale which arises due to a GUP. In the present paper our sole purpose is to work out the phase space description of such a theory. In doing so, we present the generalization of Weyl transform, Wigner function and phase space averages of operators in QM following a GUP.

As Wigner function approach to QM gives a very important insight into quantum description using phase space and is regularly used by theorists, we hope our initial attempt to formally introduce the tools required to set up a phase space picture of QM following a GUP will be important for future workers. Due to the presence of minimal length uncertainty the theory presented in this paper does not admit position eigenstates and all of the calculations are done using the momentum space. We have shown that even using the momentum space only, one can generalize the concepts of Weyl transform and Wigner function in QM following a GUP of a specific type. We have shown explicitly where the properties of the generalizations differ from the conventional properties of the quantities defined in QM following Heisenberg algebra. We explicitly calculate the Weyl transforms of ${\bf x}$ and ${\bf x}^2$ operators using our formulation and utilize these results to find out the phase space average of the quantum harmonic oscillator Hamiltonian following deformed Heisenberg algebra. Using our methods we do find out the phase space average of the harmonic oscillator energy levels and compare the results with the energy eigenvalues to show that they match remarkably. We also use our methods to obtain phase space average of a free particle analytically. These calculations shows the effectiveness of our method.

In this paper we present one of the most important results regarding
QM having a minimum length uncertainty related to the averages of
powers of various dynamical operators as ${\bf x}$, ${\bf p}$ or ${\bf
  H}$. We show that in QM with a minimal length scale, for the case of
the harmonic oscillator, phase space average values of ${\bf x}^m$,
${\bf p}^m$ and ${\bf H}^m$ for positive $m$ restrict the deformation
parameter $\beta$ which specifies the degree of deformation of
Heisenberg algebra.  We interpret this result as a genuine feature of
QM with minimal length uncertainty.

One can find the analysis for a more generalized approach in which one considers minimal uncertainties in both position and momentum (non-zero $\Delta x_0$ and $\Delta p_0$) in \cite{kempf2} - \cite{kempf5}. Particularly in \cite{kempf2}, non-commutative geometry and Bargmann-Fock algebra have been discussed and it has been examined whether this geometry could regularize the short distance behavior of quantum mechanics. In \cite{kempf3} and \cite{kempf4}, making use of the quantum group-symmetric Fock spaces with Bargmann-Fock representation, the Heisenberg Algebra is generated by the raising and lowering operators rather than the usual position and momentum operators. Here, both, bosonic and fermionic algebras are considered along with application to simple systems like driving forces. A good review of quantum mechanics in Bargmann-Fock space can be found in \cite{kempf5} along with the formulation of quantum field theory in the Fock space. The work in \cite{kempf}, and this paper, is a special case where $\Delta p_0 \rightarrow 0$.

The reason for considering this special case of only minimal position uncertainty is appearance of a minimal observable length in theories of quantum gravity. In \cite{brau}, dynamics of particle in quantum gravity well is looked into for the same GUP that we will be using in our work (as in \cite{kempf}). There are other approaches as well, for example, in \cite{jaeckel}, where limit in length measurements are associated with irreducible quantum fluctuations of geodesic distances and it is characterized by a noise spectrum with an order of magnitude mainly determined by Planck length. The non-commutative geometry, which one encounters as a result of GUP is discussed in \cite{lizzi} and \cite{maggiore}.
The assertion of minimal observable length can be found even in string theory. In \cite{konishi}, existence of minimal observable length is derived, along with the form of GUP we are about to use. Also, in \cite{amati}, it is explored whether one can probe below string size, where it is shown that it is possible as long as the energy is not large enough to cause gravitational instability. This resolution is then used to suggest a conceptual change in the meaning of spacetime.

The material in the present paper is presented as follows. In the next
section we present the basics of QM having a minimal length
uncertainty. We also introduce the conventions to write the Weyl
transform and Wigner function in standard QM following Heisenberg
algebra. In section \ref{dha} we present the formulation of Weyl
transform and Wigner function in the deformed Heisenberg algebra. In
section \ref{xx2} we present some interesting new features of the Weyl
transform in the deformed Heisenberg algebra and explicitly calculate
the Weyl transforms of ${\bf x}$ and ${\bf x}^2$ operators. In section
\ref{hod} we calculate the phase space average of the quantum harmonic
oscillator Hamiltonian in the deformed Heisenberg algebra and show how
quantum averages are affected by the use of GUP. Then in section \ref{fp}, we
consider the case of a free particle and briefly describe how its
properties are affected in presence of the GUP.  The next section
concludes the paper with a brief summary of the results.
%%%%%%%%%%%%%%%%%%%%%%%%%%%%%%%%%%%%%%%%%%%%%%%%%%%%%%%%%%%%%%%%%%%%%%%%%%%%%
\section{Deformed Heisenberg algebra and the Wigner function in standard QM}

Ideally one can deform the Heisenberg uncertainty principle (HUP) in multiple ways to obtain a GUP. In this paper we will consider the case when the GUP is such that there exists a minimal length uncertainty, $\Delta x_{0}$.
As discussed in \cite{kempf}, we assume that the GUP is of the form :
\begin{equation}
\Delta x\Delta p \geq \frac{\hbar}{2}\left(1+\beta\Delta p^{2} + \beta\langle {\bf p}\rangle ^{2}\right)\,,
\label{gupf}
\end{equation}
where $\beta>0$ and is independent of $\Delta x$ and $\Delta p$ but
in general the uncertainties can be functions of $\langle {\bf x}
\rangle$ or $\langle {\bf p} \rangle$. In this paper we will write
operators in bold letters and the eigenvalues and other functions of
c-numbers in normal font. This kind of a GUP is obtained from the
deformed Heisenberg algebra:
\begin{eqnarray}
[{\bf x}\,,\,{\bf p}]=i\hbar(1+\beta {\bf p}^2)\,,
\label{bcr}
\end{eqnarray}
which shows that $\beta$ is a dimensional constant and has the dimension of $p^{-2}$. From the GUP relation one obtains the minimal position uncertainty as:
\begin{eqnarray}
\Delta x_{\rm min}(\langle {\bf p} \rangle)=\hbar \sqrt{\beta} \sqrt{1+\beta\langle {\bf p}\rangle ^{2}}\,,
\label{dxm}
\end{eqnarray}
which gives the minimum position uncertainty to be
\begin{eqnarray}
\Delta x_{0} = \hbar \sqrt{\beta}\,.
\label{dx0}
\end{eqnarray}

Due to presence of a non-zero minimal length uncertainty, the
position eigenstates are not physical as the position uncertainty of
the position eigenstates should ideally vanish. This means, $\psi(x) = \langle
x|\psi\rangle$, loses its traditional descriptive power in QM guided
by the GUP. But as we do not consider any non vanishing minimal momentum
uncertainty, the momentum space wave function $\phi(p) = \langle
p|\psi\rangle$ still retains its descriptive power. The position and momentum operators in the momentum space description
of the deformed algebra would now look like \cite{kempf}:
\begin{equation}
{\bf x}\phi(p) = i\hbar(1+\beta p^{2})\partial_{p}\phi(p)\,,\,\,\,\,\,
{\bf p}\phi(p) = p\phi(p)\,.
\label{xprep}  
\end{equation}
Following the conventions in Ref.\cite{kempf} we can expand the identity operator in the momentum basis as
\begin{equation}
{\bf I} = \int_{-\infty}^{+\infty}\frac{dp}{1+\beta p^{2}}|p\rangle \langle p|\,,
\label{identp}  
\end{equation}
and the scalar product in the momentum basis is written as 
\begin{equation}
\langle p|p^\prime\rangle = (1+\beta p^{2})\delta (p-p^\prime)\,.
\label{spp}  
\end{equation}
These information regarding the deformed Heisenberg algebra is enough
to formulate a reasonable expression of a Wigner function in QM guided
by the GUP. Before we propose the expression of the Wigner function in
present case, we will like to specify the conventions we will follow
in defining the Wigner function. In the above analysis the magnitude
of $\beta$ is not constrained, it only depends upon the momentum scale
of the theory.  In quantum gravity theories the GUP may have
gravitational origin and the value of $\beta$ should be related to
Planck scale effects and in theories with bound states the value of
$\beta$ should be related to the momentum scale of the basic bound
states.
%%%%%%%%%%%%%%%%%%%%%%%%%%%%%%%%%%%%%%%%%%%%%%%%%%%%%%%%%%%%%%%%%%%%%%%
\subsection{Wigner Function in standard QM}
%\label{sec:level2}

Wigner Function can be defined as the quantum counterpart of classical probability distribution function. It was introduced by Eugene Wigner in 1932 \cite{wigner}. Immense work has been done on these quasi distribution \cite{case} since then and hence the literature is extensive. A few good review articles on these functions are given in \cite{hillery} and \cite{lee}. The Wigner functions can be derived from Weyl transform \cite{weyl} of operators, given by \cite{case}:
\begin{equation}
\tilde{A}(x,p) = \int e^{-\frac{ipu}{\hbar}}\langle x +u/2|{\bf A}
|x-u/2\rangle du\,,
\label{wtax}
\end{equation}
where we have used the position space description of the operator
${\bf A}$. To make the
notation less cumbersome we do not write the limits of integration
explicitly. Henceforth the integral signs appearing without any upper
or lower limits, as in the above equation, will always mean that the
integration variable runs from $-\infty$ to $+\infty$.  One can also use the momentum space description and write
the above Weyl transform as:
\begin{equation}
\tilde{A}(x,p) = \int e^{\frac{ixu}{\hbar}}\langle p+ u/2|{\bf A}
|p- u/2\rangle du\,.
\label{wtap}
\end{equation}

Although we have used the same integration
variable $u$ in both the above cases, it must be noted that $u$ is a
dimensional variable and the dimension of $u$ is not the same in the
above two equations. Writing the density operator
$\mbox{\boldmath$\rho$}$ for a pure state $|\psi \rangle$ as
$\mbox{\boldmath$\rho$}=|\psi \rangle \langle \psi |$ one defines the
Wigner Function as \cite{case}:
\begin{equation}
W(x,p) = \frac{\tilde{\rho}(x,p)}{h}
= \frac{1}{h}\int e^{-\frac{ipu}{\hbar}} \langle x+ u/2|\psi \rangle \langle \psi |x- u/2\rangle du\,,
\label{swfx}
\end{equation}
where we have used the position space representation of
$\mbox{\boldmath$\rho$}$. In this paper we will mainly discuss about
pure states, the generalization of the results for mixed states can
be done easily. In the momentum space representation of the density
operator one can write the Wigner function as
\begin{equation}
  W(x,p) = \frac{1}{h}\int e^{\frac{ixu}{\hbar}} \langle p+ u/2|\psi \rangle \langle \psi| p-u/2 \rangle du\,.
\label{swfp}
\end{equation}
In standard QM which follows HUP both the above forms of the Wigner function are equally applicable where as in QM with the deformed Heisenberg algebra the position basis is not suitable as the position operator always have a minimal, non-zero, uncertainty. In QM with GUP, Eq.~(\ref{swfp}) is more useful because it can be generalized to obtain a function which closely represents the Wigner function. In the next sections we will use an important property of Weyl transform which states that the trace of product of two operators is equivalent to the phase space integral of product of their Weyl transforms, thus \cite{case}:
\begin{equation}
{\rm Tr}[{\bf A}{\bf B}] = \frac{1}{h}\int\int\tilde{A}\tilde{B}dxdp\,.
\label{trab}
\end{equation}
In the above relation if we replace ${\bf B}$ with $\mbox{\boldmath$\rho$}$ we have \cite{case}: 
\begin{equation}
\langle {\bf A}\rangle = {\rm Tr}[\mbox{\boldmath$\rho$}{\bf A}] = \frac{1}{h}\int\int\tilde{\rho}\tilde{A}dxdp\,,
\end{equation}
which yields
\begin{equation}
\langle {\bf A}\rangle = \int\int W(x,p)\tilde{A}(x,p)dxdp\,.
\label{expa}
\end{equation}
In the next section we will show how the generalized Wigner function, in QM guided by a GUP, satisfies various properties of the standard Wigner function in QM with HUP. These various properties will convince us that we can have a phase space description of QM guided by a GUP.
%%%%%%%%%%%%%%%%%%%%%%%%%%%%%%%%%%%%%%%%%%%%%%%%%%%%%%%%%%%%%%%%%%%%%%%%%%%%%%
\section{Formulation in deformed Heisenberg algebra}
\label{dha}

The above mentioned Wigner functions are defined in normal Heisenberg
Algebra. Here, we formulate a quasi-distribution function, analogous
to the Wigner function, in the deformed Heisenberg algebra where the
uncertainties in position and momentum follows Eq.~(\ref{gupf}). We
must note that as position eigenstates do not physically exist in QM
with a minimal position uncertainty we will define the Weyl transform of an operator in the present scenario by using only the momentum basis
representation of the operator. This is a price we pay for having a minimal length uncertainty.  At first we specify how we define the Weyl transforms in the present case. The Weyl transform of any quantum mechanical operator ${\bf A}$ in the deformed Heisenberg algebra is defined as:
\begin{equation}
  \tilde{A}(x,p) \equiv \int e^{\frac{ix u}{\hbar}}\frac{\langle p+ u/2|{\bf A}|p- u/2\rangle}{\left[1+\beta (p-\frac{u}{2})^{2}\right]^{\frac{1}{2}}\left[1+\beta(p+\frac{u}{2})^{2}\right]^{\frac{1}{2}}}
  du\,.
\label{wtgup}  
\end{equation}
When $\beta \to 0$ the above definition reduces to the conventional
definition of Weyl transform of an operator. To show that the above
definition is a generalization of the Weyl transform of an operator we
first figure out the Weyl transform of the unit operator ${\bf I}$
in this formalism. The Weyl transform of the identity operator is
\begin{eqnarray}
\tilde{I} = \int e^{\frac{ix u}{\hbar}}\frac{\langle p+ u/2|{\bf I}|p- u/2 \rangle}
{\left[1+\beta (p-\frac{u}{2})^{2}\right]^{\frac{1}{2}}\left[1+\beta(p+\frac{u}{2})^{2}\right]^{\frac{1}{2}}}
du\,.\nonumber
\end{eqnarray}
Using Eq.~(\ref{spp}) we can write
$$\langle p+ u/2|p- u/2 \rangle= \left[1+\beta\left(p+\frac{u}{2}\right)^2\right]
\delta(u)\,,$$
which gives
\begin{eqnarray}
\tilde{I} = \int e^{\frac{ix
    u}{\hbar}}\frac{\left[1+\beta\left(p+\frac{u}{2}\right)^2\right]
\delta(u)}{\left[1+\beta(p-\frac{u}{2})^{2}\right]^{\frac{1}{2}}\left[1+\beta(p+\frac{u}{2})^{2}\right]^{\frac{1}{2}}}du
=1\,,
\label{wtiden}
\end{eqnarray}
as expected. Thus the form of Weyl transform assumed in
Eq.~(\ref{wtgup}) seems to work. We will verify and analyze other
properties below. In Appendix~\ref{app1} we show that our definition of the Weyl transform in QM following a GUP is unique as long as we demand the Weyl transform of the identity operator is unity and assume the validity of the basic equation for trace of operators as given in Eq.~(\ref{trab}).

We have seen in our previous discussion that the trace of product of two operators is equivalent to the phase space integral of product of their Weyl transforms as specified in Eq.~(\ref{trab}). Now we will like to show that our modified definition of the Weyl transform also satisfies this property. To prove that Eq.~(\ref{trab}) holds when QM is guided by the deformed Heisenberg algebra we first write down the integral appearing on the right hand side of Eq.~(\ref{trab}) as 
\begin{eqnarray}
\frac{1}{h}\int\int\tilde{A}\tilde{B}dxdp = \int \int \int \int \frac{e^{i\frac{(u_1 + u_2)x}{\hbar}}\langle p+ u_1/{2|{\bf A}|p- u_1/2\rangle\langle p+ u_2/2|{\bf B}|p- u_2/2\rangle}}{h\prod_{i=1,2}\left[1+\beta(p-\frac{u_i}{2})^{2}\right]^{\frac{1}{2}}\left[1+\beta(p+\frac{u_i}{2})^{2}\right]^{\frac{1}{2}}} du_1 du_2 dx dp\,.
\nonumber  
\end{eqnarray}
In our convention, the delta function is given by
$$\delta(p) = \frac{1}{2\pi \hbar}\int e^{\frac{ipx}{\hbar}} dx\,,$$ 
using which we can integrate out $x$ in the last expression as:
\begin{eqnarray}
\frac{1}{h}\int\int\tilde{A}\tilde{B}dxdp &=&\int \int \int \frac{\delta(u_1+u_2)\langle p+ u_1/2|{\bf A}|p- u_1/2\rangle\langle p+ u_2/2|{\bf B}|p- u_2/2\rangle} {\prod_{i=1,2}\left[1+\beta(p-\frac{u_i}{2})^{2}\right]^{\frac{1}{2}}\left[1+\beta(p+\frac{u_i}{2})^{2}\right]^{\frac{1}{2}}}dpdu_1du_2\nonumber\\
  &=& \int\int \frac{\langle p+ u_1/2|{\bf A}|p- u_1/2\rangle \langle p- u_1/2 |{\bf B}|p+ u_1/2 \rangle}{\left[1+\beta(p-\frac{u_1}{2})^{2}\right]\left[1+\beta(p+\frac{u_1}{2})^{2}\right]} du_1 dp\,.
\nonumber  
\end{eqnarray}
Defining new variables as $p + \frac{u_{1}}{2} = v $ and $p - \frac{u_{1}}{2} = y$
such that $dp du_1 = dvdy$ the above expression transforms to
\begin{equation}
\frac{1}{h}\int\int\tilde{A}\tilde{B}dxdp= \int\int\frac{\langle v|{\bf A}|y\rangle\langle y|{\bf B}|v\rangle}{(1+\beta y^{2})(1+\beta v^{2})}dydv\,.
\nonumber  
\end{equation}
The new variables $v$ and $y$ are effectively new momentum labels and consequently 
we can use Eq.~(\ref{identp}) to integrate out $y$ and get
\begin{equation}
\frac{1}{h}\int\int\tilde{A}\tilde{B}dxdp=\int\frac{\langle v|{\bf A} {\bf B}|v\rangle}{(1+\beta v^2)} dv\,.
\label{prel1}
\end{equation}
To make sense of the above equation we first look at the density matrix. In standard QM, for any state, the density matrix satisfies the property that
\begin{eqnarray}
{\rm Tr}[\mbox{\boldmath$\rho$}]=1\,.
\label{trr}
\end{eqnarray}
We want to keep this property unchanged in QM guided by a GUP as given in Eq.~(\ref{gupf}). In deformed Heisenberg algebra the above formula loses its full power as the trace can only be taken in the momentum basis. If we want to keep the above property of the density matrix, where Eq.~(\ref{identp}) holds, it can be seen that the trace of an operator ${\bf A}$ in momentum basis must be generalized to 
\begin{eqnarray}
{\rm Tr}[{\bf A}] \equiv \int \frac{\langle p|{\bf A}|p \rangle}{1+\beta p^2} dp\,,
\label{traced}
\end{eqnarray}
which becomes the conventional trace of an operator in momentum basis when $\beta \to 0$.
With this generalization of the trace operation, we can write Eq.~(\ref{prel1}) as
\begin{equation}
\frac{1}{h}\int\int\tilde{A}\tilde{B}dxdp= {\rm Tr}[{\bf A} {\bf B}]\,,
\nonumber
\end{equation}
producing an equation similar to Eq.~(\ref{trab}). It seems that our
definition of the Weyl transform as given in Eq.~(\ref{wtgup}) does
possess some of the known properties of Weyl transform in standard
QM. Next we define the Wigner function for a pure state $|\psi\rangle$ in the
deformed Heisenberg algebra. Writing the density operator, for a pure state $|\psi\rangle$ as,
$\mbox{\boldmath$\rho$}=|\psi\rangle \langle \psi|$ and the momentum
space wave function as $\langle p|\psi \rangle = \phi(p)$ we define
\begin{equation}
W(x,p)= \frac{\tilde\rho(x,p)}{h} \equiv \frac{1}{h}\int e^{\frac{ix u}{\hbar}}\frac{\langle p+ u/2|\psi\rangle\langle\psi|p- u/2 \rangle}{\left[1+\beta (p-\frac{u}{2})^{2}\right]^{\frac{1}{2}}\left[1+\beta(p+\frac{u}{2})^{2}\right]^{\frac{1}{2}}}du\,,
\label{wfgup}  
\end{equation}
which can also be written as
\begin{equation}
W(x,p) = \frac{1}{h}\int e^{\frac{ix u}{\hbar}} \frac{\phi(p+ u/2)\phi ^*(p- u/2)}{\left[1+\beta (p-\frac{u}{2})^{2}\right]^{\frac{1}{2}}\left[1+\beta(p+\frac{u}{2})^{2}\right]^{\frac{1}{2}}}du\,.
\label{wfgup1}
\end{equation}
In this paper we focus on the pure states, the mixed state results can
be easily obtained once the basic pure state results are known. It is
trivial to show that the Wigner function defined above is a real
function. As Eq.~(\ref{trab}) remains the same in the present case, it
predicts that the expectation value of any operator ${\bf A}$ in the
deformed Heisenberg algebra can still be written as
$$\langle {\bf A}\rangle = {\rm Tr}[{\mbox{\boldmath$\rho$}}{\bf A} ] = \int\int W(x,p)\tilde A(x,p)dxdp\,,$$ where
$W(x,p)$ is given by Eq.~(\ref{wfgup1}) and $\tilde A(x,p)$ is given
by Eq.~(\ref{wfgup}). This observation is very important as it enables
us to find out the expectation values of various operators in QM
guided by the GUP. The above fact establishes that in QM guided by the specified GUP, the Wigner function still acts like a quasi-distribution function which can be used to find out the phase space average of an operator. 

Recently some authors have tried to address the the difficulties arising in theories with GUP in classical mechanics as well as in QM \cite{Bosso:2018uus, Bosso:2020aqm}. The difficulties are related with the fact that the position and momentum operators we are working with in this paper are not conjugate to each other. Properly conjugate variables must follow the standard Heisenberg algebra. As because the position and momentum variables are not conjugate to each other ambiguities can arise when transforming from the Lagrangian to the Hamiltonian formulation. The aforementioned works then devise a new set of canonical variables as $({\bf p}_0, {\bf x}_0)$ which are related to the physical momentum and coordinate operators $({\bf p},{\bf x})$.  The new operators $({\bf p}_0, {\bf x}_0)$ satisfies standard Heisenberg algebra. These two sets of variables are functionally related. We bring up this point here as it may be interesting to see whether these transformations can help us to formulate the Wigner function in terms of the new variables $({\bf p}_0, {\bf x}_0)$. One may expect that such a Wigner function will be easy to handle  as the phase space variables with which it is defined are now properly conjugate variables. In this regard our observation is that it is indeed interesting to do such an exercise but at the end of it we have to come back, or map the Wigner function back, to the physical non-conjugate variables $({\bf p},{\bf x})$. This mapping by itself can be a new project which we do not
discuss in this paper. In this paper we have avoided the use of the auxiliary canonical conjugate variables $({\bf p}_0, {\bf x}_0)$ and have directly worked with the non-conjugate variables $({\bf p},{\bf x})$.

Before we finish our discussion on the definition of Wigner function, in QM following modified Heisenberg algebra, we will like to inform the reader that there exists a different formulation for the Wigner function in modified Heisenberg algebra as specified in Ref.~\cite{Das:2014bba}. In the previous attempt the authors did not modify the definition of the Weyl transform in modified Heisenberg algebra and as a result they used the expression of Wigner function as given in Eq.~(\ref{swfp}) with GUP modified wavefunctions. As a consequence their formulation looks simple compared to the present work but the outcome of their formulation produces different fundamental  results. To see how the previous results can be compared with the present work we first calculate the Weyl transform of the identity operator in their formalism. In their formalism
\begin{equation}
\tilde{I} = \int e^{\frac{ixu}{\hbar}}\langle p+u/2|\textbf{I}|p-u/2\rangle\,du\,.
\end{equation}
As we know that in the theory of GUP we are working with we have
\begin{equation}
\langle p+u/2|p-u/2\rangle = [1+\beta(p+u/2)^2]\delta(u)\,,
\end{equation}
the Weyl transform of the identity operator becomes
\begin{equation}
\tilde{I} = \int e^{\frac{ixu}{\hbar}}[1+\beta(p+u/2)^2]\delta(u)du = (1+\beta p^2)\,.
\end{equation}
It shows that if one does not modify the Weyl transform appropriately in GUP modified QM one does not get unity as the Weyl transform of the identity operator. One has to rescale the above identity operator with a momentum dependent factor to obtain unity.

Next we will try to see whether the Wigner function as defined in Ref.~\cite{Das:2014bba} can be taken as a quasi phase space distribution in GUP modified QM. Following a similar calculation presented previously we first notice that 
\begin{equation}
\frac{1}{h}\int\int\tilde{A}\tilde{B}dxdp 
=\int \int \langle p+u_1/2|\textbf{A}|p-u_1/2\rangle\langle p-u_1/2|\textbf{B}|p+u_1/2\rangle \,dp\,du_1\,.
\end{equation}
Substituting $p+u_1/2=v$, $p-u_1/2=y$ we get
\begin{equation}
\frac{1}{h}\int\int\tilde{A}\tilde{B}dxdp = \int \langle v|\textbf{A}|y\rangle\langle y|\textbf{B}|v\rangle dvdy\,.
\end{equation}
Next we write the trace of the product of the operators ${\bf A}$ and ${\bf B}$ as:
\begin{equation}
{\rm Tr}[\textbf{AB}] = \int \frac{\langle v|\textbf{AB}|v\rangle}{(1+\beta v^2)}dv = \int \frac{\langle v|\textbf{A}|y\rangle\langle y|\textbf{B}|v\rangle}{(1+\beta v^2)(1+\beta y^2)}\,dv\,dy
\end{equation}
where we have used the modified definition of trace and the
completeness relation in Eq.~(\ref{identp}). Comparing these
previous results we see
\begin{equation}
{\rm Tr}[\textbf{AB}] \ne \frac{1}{h}\int \tilde{A}\tilde{B}\,dx\,dp\,,
\end{equation}
in the formalism developed in Ref.~\cite{Das:2014bba}. If one does not
use the modified definition of trace, follows the conventional
definition, but only uses the modified completeness relation then also
the above inequality holds true. Due to the above inequality we have
\begin{equation}
\langle{\bf A}\rangle={\rm Tr}[{\bf \rho}{\bf A}] \ne \frac{1}{h}\int\int\tilde{\rho}\tilde{A}dxdp\ = \int\int W(x,p)\tilde{A}(x,p)dxdp,\,
\end{equation}
and as a result the Wigner function as defined in the referred work does not act like a quasi phase space distribution function. These examples show that if one does not modify the definition of the Weyl transform in QM modified by a GUP, as done in the present paper, then the Wigner function cannot be used in the most general sense as a quasi-distribution function.
The present work logically completes the interesting discussion initiated in Ref.~\cite{Das:2014bba}.
%%%%%%%%%%%%%%%%%%%%%%%%%%%%%%%%%%%%%%%%%%%%%%%%%%%%%%%%%%%%%%%%%%%%%%%%%%%%%%%%%%%%%%
\subsection{Properties of the Quasi-distribution function in QM following GUP}

We present some of the properties of the Wigner function in QM guided by the GUP as given in Eq.~(\ref{gupf}). In the present case we have a minimal length uncertainty $\Delta x_{0}$
as a result of which position eigenkets cannot be defined. Suppose we are working with a normalized state $|\psi\rangle$ such that $\langle \psi|\psi \rangle=1$, in such a case we can show that the generalized Wigner function as given in Eq.~(\ref{wfgup}) is normalized to unity. To show this result we multiply both sides of Eq.~(\ref{wfgup1}) by $dx$ and $dp$ and then integrate to get 
\begin{equation}
\int\int W(x,p)dx dp = \frac1h\int \int \int \frac{e^{ix u/\hbar}\phi(p+ u/2)\phi^*(p- u/2)}
{\left[1+\beta (p-\frac{u}{2})^{2}\right]^{\frac{1}{2}}\left[1+\beta(p+\frac{u}{2})^{2}\right]^{\frac{1}{2}}}
du dx dp\,.
\nonumber
\end{equation}
We can now integrate out $x$ using the Dirac delta function defined earlier and get 
\begin{eqnarray}
\int\int W(x,p)dx dp &=& \frac{2\pi\hbar}{h}\int \frac{\delta(u)\phi(p+ u/2)\phi^*(p- u/2)}{\left[1+\beta (p-\frac{u}{2})^{2}\right]^{\frac{1}{2}}\left[1+\beta(p+\frac{u}{2})^{2}\right]^{\frac{1}{2}}} du dp\nonumber\\
&=& \int \frac{\phi(p)\phi^{*}(p)}{(1+\beta p^{2})}dp
= \int \frac{\langle\psi|p\rangle\langle p|\psi\rangle}{(1+\beta p^{2})}dp\,.
\nonumber
\end{eqnarray}
Now using the expression of the identity operator as given in Eq.~(\ref{identp}) the above equation becomes
\begin{equation}
\int\int W(x,p)dx dp = \langle\psi|\psi\rangle = 1\,,
\label{normwf}
\end{equation}
showing that the Wigner function is normalized in the phase space.

Next we show that in QM guided by a GUP as given in Eq.~(\ref{gupf}) the 
integral of the square of the Wigner Function in phase space is bounded as it happens in standard QM.
For a pure state $|\psi \rangle$ which is normalized to unity we know $\mbox{\boldmath$\rho$}^2=\mbox{\boldmath$\rho$}$ and consequently
\begin{equation}
{\rm Tr}[\mbox{\boldmath$\rho$}^2]={\rm Tr}[\mbox{\boldmath$\rho$}] = 1\,,
\label{rsqr}  
\end{equation}
where we have used the result of Eq.~(\ref{trr}). Using Eq.~(\ref{trab}), which holds in the present case also, we can now write
\begin{equation}
\frac{1}{h}\int\int\tilde\rho^2(x,p) dp dx= 1\,,
\nonumber  
\end{equation}
which can also be written as:
\begin{equation}
\int\int W^2(x,p)dpdx = \frac{1}{h}\,,
\label{bwfg}  
\end{equation}
using the definition of the Wigner function. Because of the above relation we see that the square of the Wigner function still remains bounded in the phase space.  

Suppose $|\psi_A\rangle$ and $|\psi_B \rangle$ be two states in QM following the deformed Heisenberg algebra. In such a case we can always write
\begin{eqnarray}
{\rm Tr}[\mbox{\boldmath$\rho$}_A\mbox{\boldmath$\rho$}_B]&=&h^{-1}\int \frac{\langle p|\psi_A\rangle \langle  \psi_A|\psi_B \rangle \langle \psi_B|p\rangle}{1+\beta p^2} dp\nonumber\\
&=& h^{-1}\int \frac{\langle  \psi_A|\psi_B \rangle \langle \psi_B|p\rangle \langle p|\psi_A\rangle }{1+\beta p^2} dp\nonumber\\
&=& h^{-1}|\langle  \psi_A|\psi_B \rangle|^2\,,
\label{trrarb}
\end{eqnarray}
where we have used the expression of identity as given in Eq.~(\ref{identp}). In Ref.~\cite{kempf} it has been discussed that wave-functions $\phi(p)$ in the deformed Heisenberg algebra can be normalized with respect to the measure $dp/(1+\beta p^2)$, which implies
\begin{eqnarray}
\langle  \psi_A|\psi_B \rangle=\int \frac{\phi_B(p)\phi_A^*(p)}{1+\beta p^2} dp\,,
\label{sprod}
\end{eqnarray}
where $\phi_A(p) = \langle p|\psi_A\rangle$ and $\phi_B(p) = \langle p|\psi_B \rangle$. Using the relation in Eq.~(\ref{trab}) we can now write Eq.~(\ref{trrarb}) as
\begin{equation}
\int\int W_A(x,p) W_B(x,p)dx dp=h^{-1}|\langle\psi_A|\psi_B\rangle|^{2}\,.
\end{equation}
The above equation will give $0$ for A$\neq$ B. So, the Wigner functions in our new formulation also satisfy the properties of its old normal quantum mechanics counterpart. For orthogonal states where
$\langle  \psi_A|\psi_B \rangle=0$ we must have
$$\int\int W_A(x,p) W_B(x,p)dx dp=0\,,$$
showing that like the standard result, the Wigner function in the deformed Heisenberg algebra must have negative values in the phase space.

If $|\psi_2 \rangle$ is a normalized state in QM following GUP then from Eq.~(\ref{sprod}) we have
\begin{eqnarray}
\langle  \psi_2|\psi_2 \rangle=\int \frac{|\phi_2(u)|^2}{1+\beta u^2} du=1\,.
\label{sprod1}
\end{eqnarray}
From this equation one can easily verify that 
$$\phi_2(u)=\left(\frac{1+\beta u^2}{2}\right)^{1/2}\frac{\phi(p-u/2)}{\left[1+\beta(p-\frac{u}{2})^{2}\right]^{\frac{1}{2}}}\,,$$
as well as
$$\phi_1(u)=e^{ixu/\hbar}\left(\frac{1+\beta
  u^2}{2}\right)^{1/2}\frac{\phi(p+u/2)}{\left[1+\beta(p+\frac{u}{2})^{2}\right]^{\frac{1}{2}}}\,,$$
are normalized momentum space wave functions in QM following GUP. In
these above equations $p$ specifies a constant momentum eigenvalue.
Using these normalized wave functions we can now write the expression
of the Wigner function as given in Eq.~(\ref{wfgup1}) as:
\begin{eqnarray}
  W(x,p) = \frac{2}{h}\int \frac{\phi_1(u)\phi^*_2(u)}{1+\beta u^2}\,du = \frac{2}{h}\langle \psi_2|\psi_1\rangle\,.
\label{wfbn}  
\end{eqnarray}
This above calculation implies
\begin{eqnarray}
|W(x,p)| \le \frac2h\,,
\label{wfbnm}  
\end{eqnarray}
showing that the Wigner function is bounded from above in QM modified by GUP. The upper bound remains exactly the same as in standard QM following Heisenberg algebra.  

In this section we saw that the Wigner function as defined in QM
following a deformed Heisenberg algebra has many of its properties
similar to the conventional Wigner function in standard QM following
normal Heisenberg algebra. Before we end this discussion
we would like to point out that the Wigner function
presented in this section does have some differences with the
conventional function. The Wigner function expression presented in the
above discussion lacks any translation property in coordinate
representation. It is known that in conventional QM if $\psi(x)$
produces the Wigner function $W(x,p)$ then $\psi(x-x_0)$ will produce
a Wigner function $W(x-x_0,p)$. In the present case as we do not work
with position wave functions this translation property is not
apparent. If the momentum wave function $\phi(p)$
  gives us the Wigner function $W(x, p)$, then we can easily see that
  $\phi(p-p_0)$ will not give us $W(x, p-p_0)$ due to the factors
  present in the denominator which become equal to unity only when one
  goes to standard quantum mechanics by putting $\beta = 0$. Thus,
  shifts in the wave function do not lead to corresponding shifts in
  the Wigner function in the momentum eigenvalue $p$. When we replace
  the original wave function with $\phi(p)e^{-ix_0p/\hbar}$, the new
  Wigner function will become $W(x-x_0, p)$. Thus the wave functions
  of $W(x, p)$ and $W(x-x_0, p)$ differ by a phase $e^{-ix_0p/\hbar}$
As a consequence we can summarize that the Wigner function, as defined
in this section, for the deformed Heisenberg algebra can reasonably be
accepted as a phase space quasi-distribution. In the next section we will point out some important new
features of Weyl transforms arising out of the deformed Heisenberg
algebra.
%%%%%%%%%%%%%%%%%%%%%%%%%%%%%%%%%%%%%%%%%%%%%%%%%%%%%%%%%%%%%%%%%%%%%%%%%%%%%%%%%%%%%%%%%%%
\section{New features of Weyl transform in the deformed Heisenberg algebra}
\label{xx2}

Despite the similarity of the new Wigner function with the
conventional one, the new quasi distribution formulation does lack
some important features.  Earlier, if an operator ${\bf A}$ is purely
a function of ${\bf x}$, then its Weyl transform is just the original
function with the operator ${\bf x}$ replaced by position variable
$x$. If we start with an operator dependent only on the momentum
operator {\bf p}, we find a similar result. In the deformed Heisenberg
Algebra, we do not have any position eigenstate, so if ${\bf A} =
f({\bf p})$, where $f$ represents an arbitrary well behaved function,
we will have $\tilde{A} = f(p)$. But we cannot conclude similarly for
a position operator function, i.e. if ${\bf A} = f({\bf x})$ then
$\tilde{A} = f(x)$ will not hold. This is due to the fact that we do
not have an equivalent definition of distribution function in position
eigenspace which was possible in conventional QM, therefore we do not
have a simple Weyl transform of an ${\bf x}$ dependent operator. In
this section we find out the Weyl transforms of ${\bf x}$ and ${\bf
  x}^2$ operators for later use.
%%%%%%%%%%%%%%%%%%%%%%%%%%%%%%%%%%%%%%%%%%%%%%%%%%%%%%%%%%%%%%%%%%%%%%%%%%%%
\subsection{Weyl transform of ${\bf x}$}

In the deformed Heisenberg algebra we have
\begin{equation}
\tilde{x}=\int e^{\frac{ix u}{\hbar}}\frac{\langle p+ u/2|{\bf x}|p- u/2\rangle}{\left[1+\beta (p-\frac{u}{2})^{2}\right]^{\frac{1}{2}}\left[1+\beta(p+\frac{u}{2})^{2}\right]^{\frac{1}{2}}}du\,.
\label{xtld}  
\end{equation}
The {\bf x} operator in the momentum basis is given as
${\bf x} = i\hbar(1+\beta p^2) \partial_p$ and as a consequence we can write
\begin{eqnarray}
\langle p+ u/2|{\bf x}|p- u/2\rangle
= i\hbar g(u,p)\frac{d}{du}\delta(u)\,,
\label{xpbs}
\end{eqnarray}
where $g(u,p)=[1+\beta(p+u/2)^2][1+\beta (p-u/2)^2]$. The Weyl transform of ${\bf x}$ comes out as
\begin{equation}
\tilde{x}=i\hbar\int du \,\, e^{\frac{ixu}{\hbar}}\sqrt{g(u,p)}\frac{d}{du}\delta (u)\,.
\end{equation}
The integral yields
\begin{equation}
\tilde{x}=x(1+\beta p^2)
\label{tilx}  
\end{equation}
It is easily observable from the above equation that in the limit $\beta=0$ we recover our expected normal quantum mechanics result. The above result shows that in the deformed Heisenberg  algebra the Weyl transform of position operator is momentum dependent. This result is an important one showing that although in the present case the position basis ideally does not exist we can still define the Weyl transform of the position operator.  In standard QM following Heisenberg algebra the Weyl transform of the position operator should have a sharp value and that coincides with the position eigenvalue. In the present case the value of the Weyl transform of position operator is modified by the square of the momentum eigenvalue and the magnitude of the deformation parameter $\beta$. Although the definition of the Weyl transform of an arbitrary operator function does depend upon an unique position marker $x$, through the exponential function in Eq.~(\ref{wtgup}), the Weyl transform of the position operator is modified by the  the momentum eigenvalue.  This result indirectly shows that we are dealing with a case where sharp position coordinates ideally do not exist.  In the next subsection we calculate the Weyl transform of ${\bf x}^2$ operator.
%%%%%%%%%%%%%%%%%%%%%%%%%%%%%%%%%%%%%%%%%%%%%%%%%%%%%%%%%%%%%%%%%%%%%%%%%%%%%%%%%%%%%%%%
\subsection{Weyl transform of ${\bf x}^2$}

The operator ${\bf x}^2$ is:
\begin{equation}
{\bf x}^2= -\hbar^{2}(1+\beta p^{2})^2 \partial_p^{2}- 2\beta\hbar^{2}(1+\beta p^{2})p\partial_p\,,
\end{equation}
and 
\begin{equation}
\tilde{x}^2= \int e^{\frac{ix u}{\hbar}}\frac{\langle p+ u/2|{\bf x}^2|p- u/2\rangle }{\left[1+\beta (p-\frac{u}{2})^{2}\right]^{\frac{1}{2}}\left[1+\beta(p+\frac{u}{2})^{2}\right]^{\frac{1}{2}}}du\,.
\label{xtsq}
\end{equation}
To evaluate the non-exponential term in the numerator of the integrand we use 
$$\langle \xi_1|{\bf x}^2|\xi_2 \rangle =\int \frac{\xi_1(p^\prime){\bf x}^2 \xi_2(p^\prime)}{1+\beta p^{\prime 2}}\,dp^\prime\,,$$
where $\xi_1(p^\prime)=\left[1+\beta\left(p+\frac{u}{2}\right)^2\right]\delta(p^\prime - p - \frac{u}{2})$ and $\xi_2(p^\prime)=\left[1+\beta\left(p-\frac{u}{2}\right)^2\right]\delta(p^\prime - p + \frac{u}{2})$. Once we plug in the form of ${\bf x}^2$ in the above integral the individual terms in the above integral turns out to be
\begin{equation}
\langle p+ u/2|-\hbar^2(1+\beta p^{\prime 2})^2\partial_{p^{\prime}}^2|p- u/2\rangle 
= -g(u,p)\hbar^{2}\left[1+\beta \left(p+\frac{u}{2}\right)^2\right]\frac{d^2}{du^2}\delta (u)\,,
\label{xtsq1}
\end{equation}
and 
\begin{equation}
\langle p+ u/2|-2\beta p^\prime\hbar^{2}(1+\beta {p^\prime}^2)\partial_{p^\prime}|p- u/2 \rangle = -g(u,p)2 \hbar^2 \beta \left(p+\frac{u}{2}\right)\frac{d}{du}\delta (u)\,, 
\label{xtsq2}
\end{equation}
where $g(u,p)=[1+\beta(p+u/2)^2][1+\beta (p-u/2)^2]$ as defined previously.
Using the expression in Eq.~(\ref{xtsq1}) we have
\begin{eqnarray}
-\hbar^{2}\int e^{\frac{ix u}{\hbar}}\frac{g(u,p)\left[1+\beta(p+\frac{u}{2})^{2}\right]\delta^{\prime\prime}(u)}{\left[1+\beta (p-\frac{u}{2})^{2}\right]^{\frac{1}{2}}\left[1+\beta(p+\frac{u}{2})^{2}\right]^{\frac{1}{2}}}du
&=& x^2(1+\beta p^2)^2-2i\hbar x\beta p(1+\beta p^2)\nonumber\\
&-&\beta \hbar^2(1+\beta p^2)+\hbar^2\beta^2 p^2\,.
\nonumber
\end{eqnarray}
Next using the expression in Eq.~(\ref{xtsq2}) we can write
\begin{equation}
-\hbar^{2}\int e^{\frac{ix u}{\hbar}}\frac{2 g(u,p)\beta\left(p+\frac{u}{2}\right)\delta^{\prime}(u)}{\left[1+\beta (p-\frac{u}{2})^{2}\right]^{\frac{1}{2}}\left[1+\beta(p+\frac{u}{2})^{2}\right]^{\frac{1}{2}}}du
=\beta \hbar^2(1+\beta p^2)+2i\hbar \beta px (1+\beta p^2)\,.
\nonumber
\end{equation}
From the last two equations we obtain the Weyl transform of ${\bf x}^2$ as
\begin{equation}
\tilde{x}^2=x^2(1+\beta p^2)^2+\hbar^2\beta^2p^2\,.
\label{wtxsqr}  
\end{equation}
It is seen that the Weyl transform of ${\bf x}^2$ is not exactly the square of the Weyl transform of ${\bf x}$, the two Weyl transforms differ by a factor proportional to $\hbar^2$. 
We can easily see that in the limit $\beta \rightarrow 0$, we obtain the normal quantum mechanical result. The above results shows the nature of the Weyl transform operation in deformed Heisenberg algebra and the ways this transform differs from the results in conventional QM. Next we will take the concrete case of the linear harmonic oscillator, whose quantization follows the rules of the deformed Heisenberg algebra, and present its phase space properties.
%%%%%%%%%%%%%%%%%%%%%%%%%%%%%%%%%%%%%%%%%%%%%%%%%%%%%%%%%%%%%%%%%%%%%%%%%%%%%%%%%%%%%%%%%%%%%%
\section{Harmonic Oscillator in Deformed Heisenberg Algebra}
\label{hod}

Quantum harmonic oscillators have been studied in the context of GUP in Ref.~\cite{Bosso:2017ndq}. In the work mentioned the authors have used ladder operators to address various issues in GUP modified quantum mechanics of harmonic oscillators. A general technique to use ladder operators in QM was devised in \cite{Bosso:2018syo} which can be useful in GUP modified theories as well. In this paper we will not use ladder operators to evaluate the Wigner function in QM modified by GUP. We will use the wave functions obtained by solution of the Schroedinger equation. The Hamiltonian of the harmonic oscillator is given by
\begin{equation}
{\bf H} = \frac{{\bf p}^2}{2m} + m\omega^2\frac{{\bf x}^2}{2}\,,
\end{equation}
and we assume that the the quantum theory based on this Hamiltonian satisfies the deformed Heisenberg algebra as given in Eq.~(\ref{bcr}). The wave functions for the harmonic oscillator in deformed Heisenberg algebra are rigorously derived in Kempf et al. (1995) and are given by \cite{kempf}:
\begin{equation}
\phi_{n}(p) \propto \frac{1}{(1+\beta p^{2})^{\sqrt{q+r_{n}}}}\,F\left(a_{n},-n,c_{n};\frac{1+i\sqrt{\beta}p}{2}\right)\,,
\label{wfho}  
\end{equation}
where $F(a,b,c,z)$ is the Hypergeometric function. The energy levels are given by \cite{kempf}:
\begin{equation}
  E_{n} = \hbar\omega\left(n+\frac{1}{2}\right)\left(\frac{1}{4\sqrt{r}}+\sqrt{1+\frac{1}{16r}}\right)
+\frac{\hbar\omega n^{2}}{4\sqrt{r}}\,,
\label{eev}  
\end{equation}
where 
\begin{equation}
\eta^{2} = \frac{1}{(m\hbar \omega)^{2}}\,,\,\,\,\,\,
r = \frac{\eta^{2}}{4\beta^{2}}
\label{r}
\end{equation}
and
\begin{equation}
 a_{n} = -n-\sqrt{1+16r}\,,\,\,\,\,\,\,\,\, \sqrt{q+r_{n}} = \frac{1}{2}\left(n+\frac{1}{2}\right)+\frac{1}{4}\sqrt{1+16r}\,,\,\,\,\,\,\,\,\,
 c_{n} = 1-2\sqrt{q+r_{n}}\,.
\nonumber 
\end{equation}
From the above equations, we can write the wave functions for various values of $n$ which will be used for Wigner function construction, as  
\begin{equation}
    \phi_{0}(p)\propto\frac{1}{(1+\beta p^2)^{0.25+\sqrt{5}/4}}\,,\,\,\,\,\,
    \phi_{1}(p)\propto\frac{-ip}{(1+\beta p^2)^{0.75+\sqrt{5}/4}}\,,\,\,\,\,
    \phi_{2}(p)\propto\frac{1-(2+\sqrt{5})p^2}{(3+\sqrt{5})(1+\beta p^2)^{1.25+\sqrt{5}/4}}
\nonumber
\end{equation}
and

\begin{equation}
\phi_{3}(p)\propto\frac{i\left[(2+\sqrt{5})p^3-3p\right]}{(5+\sqrt{5})(1+\beta p^2)^{1.75+\sqrt{5}/4}}\,,
\nonumber  
\end{equation}
where the proportionality sign is later turned into equality by use of the normalization of the wave functions
\begin{equation}
\int\frac{|\phi(p)|^2}{(1+\beta p^2)}dp=1\,,
\label{pnorm} 
\end{equation}
arising from the definition of the scalar product of wave-functions given in Eq.~(\ref{sprod}). For the phase space description of QM, in the present case, we will plot the shape of the Wigner function as introduced via Eq.~(\ref{wfgup1}).  The Wigner functions for the various cases will be obtained numerically. In the present case, the Weyl transform of the Hamiltonian is given by 
\begin{equation}
    \tilde{H}(x,p) = \int e^{\frac{ix u}{\hbar}}\frac{\langle p+ u/2|{\bf H}|p- u/2\rangle}{\left[1+\beta (p-\frac{u}{2})^{2}\right]^{\frac{1}{2}}\left[1+\beta(p+\frac{u}{2})^{2}\right]^{\frac{1}{2}}}du
= \frac{\tilde p^{2}}{2m} + m\omega^{2}\frac{\tilde x^{2}}{2m}\,,
\nonumber
\end{equation}
which becomes
\begin{equation}
\tilde{H}(x,p) = \frac{p^2}{2m} + \frac{x^2(1+\beta p^2)^2+\hbar^2\beta^2p^2}{2}\,,
\label{lhoha}
\end{equation}
following the results from the last section, in particular Eq.~(\ref{wtxsqr}). Using the Wigner function corresponding to various $n$ values we can, in principle, find out the average value of ${\bf H}$. The average value of the Hamiltonian is
\begin{equation}
  \langle {\bf H}\rangle=\int\int W(x,p)\left[\frac{p^2}{2m} + \frac{x^2(1+\beta p^2)^2+\hbar^2\beta^2p^2}{2}\right] dxdp\,.
\label{avh}  
\end{equation}
We expect $\langle {\bf H}\rangle$ calculated in the phase space to be equal to the energy eigenvalue give in Eq.~(\ref{eev}) for any particular $n$. In this section we show that we do get the expected result with our modified definition of the Wigner function. 

We numerically computed the phase space average of the Hamiltonian
when the system was in ground state, 1st, 2nd and 3rd excited states using Eq.~(\ref{avh})
and compared them to the values calculated by using the analytical
formula of energy eigenvalues, as given in Eq.~(\ref{eev}). In our
calculation we took the values of m, $\hbar$, $\omega$ to be 1. We
want to check whether the two formulations are mathematically
consistent with each other, so any arbitrary value of $\beta$ should
yield same value of energy levels for both. Hence, for simplicity we
take $\beta=1$ and simulate.  These values are tabulated below in Table 1.

%%%%%%%%%%%%%%%%%%%%%%%%%%%%%%%%%%%%%%%%%%%%%%%%%%%%%%%%%%%%%%%%%%%%%%%%%%%%%%%%%%%
\begin{table}[h]
\begin{center}
\begin{tabular}{crl}
\hline\hline
\multicolumn{1}{c}{Energy Level} &
\multicolumn{1}{c}{Energy eigenvalue} &
\multicolumn{1}{c}{Phase space average}\\
\multicolumn{1}{c}{} &
\multicolumn{1}{c}{}  \\ \hline
$n=0$ &  $0.80901$ \hspace{2mm}& $0.80685$  \\
$n=1$&$2.92705$ \hspace{2mm}&
$2.91718$ \\
$n=2$&$6.04508$ \hspace{2mm}&
$6.01935$\\
$n=3$&$10.16312$ \hspace{2mm}& $10.11012$ \\
\hline\hline
\end{tabular}
\caption{Energy levels calculated from eigenvalues and phase space average.}
\label{tab:experiments}
\end{center}
\end{table}
\vspace{0cm}
%%%%%%%%%%%%%%%%%%%%%%%%%%%%%%%%%%%%%%%%%%%%%%%%%%%%%%%%%%%%%%%%%%%%%%%%%%%%%%%%%%%%%

The table content shows that there is reasonable agreement between the analytically calculated eigenvalues of energy and the average energy values of the oscillator following deformed Heisenberg algebra. The slight mismatch is due to numerical error encountered while performing the phase space integrals. This explicit calculation shows that our prescription for the Wigner function in QM following deformed Heisenberg algebra can be used for meaningful purpose. 

\begin{figure}[h!]
%\centering
%\begin{minipage}{.48\textwidth}
  \center
  \includegraphics[width=1.2\linewidth]{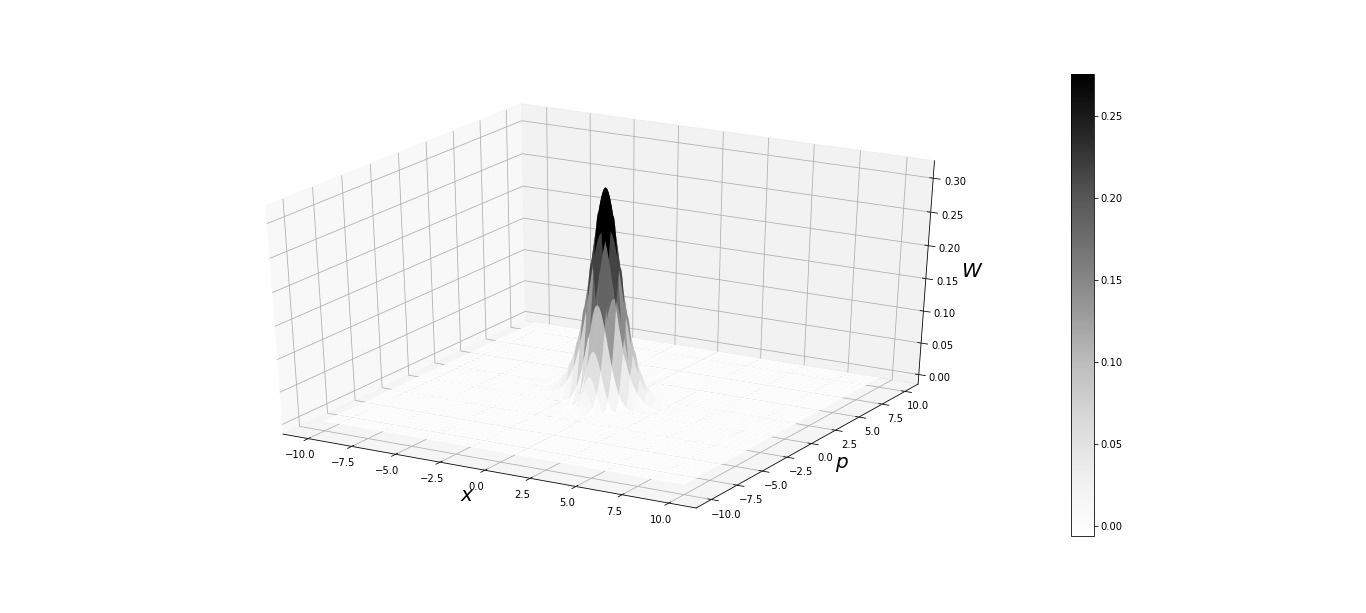}
  \caption{Wigner function for the harmonic oscillator in QM following GUP for $n=0$. Details regarding the parameters used to plot the graph can be found in text.}
  \label{fig:test1}
%\end{minipage}%
%\begin{minipage}{.48\textwidth}
%  \centering
%  \includegraphics[width=1.5\linewidth]{n1_topright.png}
%  \caption{Wigner function for n=1}
%  \label{fig:test2}
%\end{minipage}
\end{figure}

\begin{figure}[h!]
%\centering
%\begin{minipage}{.48\textwidth}
  \center
  \includegraphics[width=1.2\linewidth]{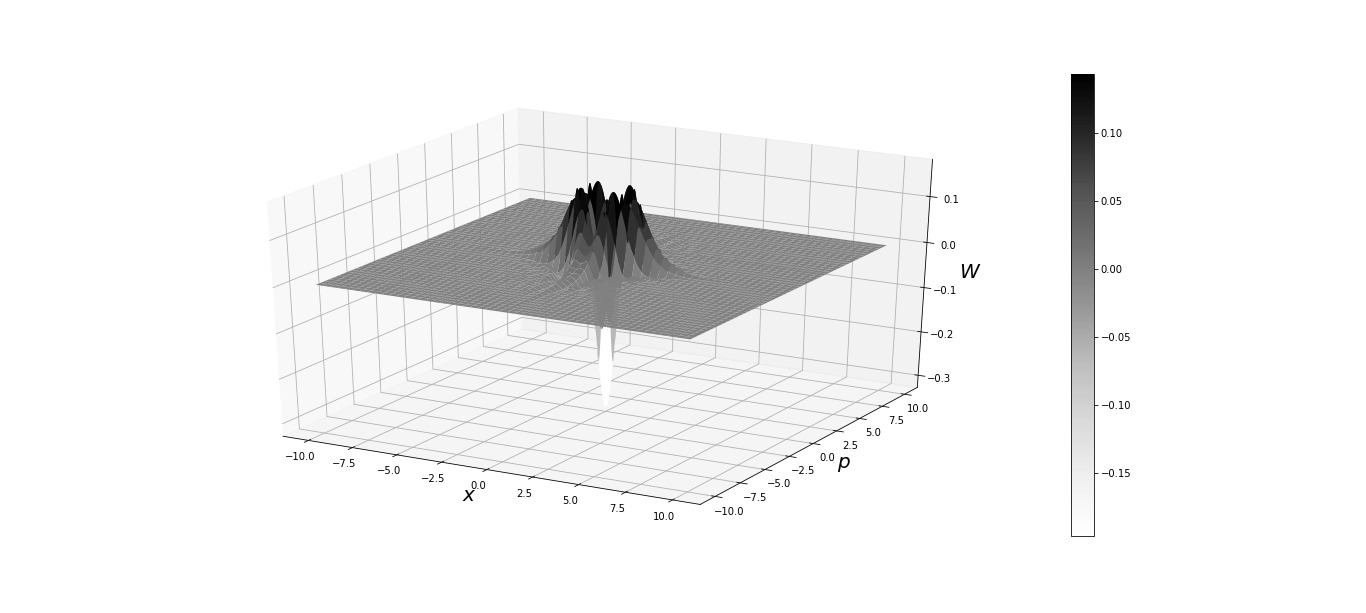}
  \caption{Wigner function of the harmonic oscillator in QM following GUP for the state $n=1$. Details regarding the parameter choices used in plotting the graph is given in text.}
  \label{fig:test2}
%\end{minipage}%
%\begin{minipage}{.48\textwidth}
%  \centering
%  \includegraphics[width=1.5\linewidth]{n1_topright.png}
%  \caption{Wigner function for n=1}
%  \label{fig:test2}
%\end{minipage}
\end{figure}

The plots in Fig.~\ref{fig:test1} and Fig.~\ref{fig:test2} show the Wigner function of the Harmonic oscillator following GUP for $n=0$ and $n=1$. We present these two plots to show how the Wigner function in the present case looks like. The shapes of the plots are not too different from the shape of the Wigner functions for the conventional harmonic oscillator in QM following Heisenberg algebra. The parameter values used to plot the above curves were discussed in the previous paragraph. 
%%%%%%%%%%%%%%%%%%%%%%%%%%%%%%%%%%%%%%%%%%%%%%%%%%%%%%%

Till now we have shown that we can have a phase space picture in QM
following a GUP. Next we will discuss how the phase space picture
points out towards a vital limitation of the theory. As the GUP
deforms the phase space volume, and this deformation is dependent on
the expectation value of the momentum operator, we expect the phase
space averages to be affected by the deformation of phase space in the
high momentum limit. The deformation of phase space is specified by
the parameter $\beta$ in the present case. We expect the phase space
averages to be dependent on $\beta$, though this dependence will vary as we change the system, we will now show that the phase
space averages for operators like ${\bf x}^m$, ${\bf p}^m$ and ${\bf
  H}^m$, where $m$ is some positive integer, produce a limit on the
possible value of $\beta$.
%%%%%%%%%%%%%%%%%%%%%%%%%%%%%%%%%%%%%%%%%%%%%%%%%%%%%%%%%%%%%%%%%%%%%%%%
\subsection{The bound on $\beta$ arising from averages}

For the Wigner function specifying the various states of the harmonic
oscillator we will use the expression as given in
Eq.~(\ref{wfgup1}). In our case the the momentum space wave functions,
for the harmonic oscillator in the deformed Heisenberg algebra, are
specified at the beginning of this section. The expectation value of
any operator expressed as a phase space average is given by $\langle
{\bf A}\rangle = \int\int W(x,p)\tilde A(x,p)dxdp,$ where the Weyl
transform $\tilde A(x,p)$, in the deformed Heisenberg algebra, is
given in Eq.~(\ref{wtgup}).  Let us concentrate on the power of
momentum $p$ in the integrand of the phase space integration giving
$\langle {\bf A}\rangle$. In the high momentum limit we must have
$W(x,p) \propto p^{2(n-2j)-2}$ where $n$ is the order of the
polynomial function and
$$j \equiv
\sqrt{q+r_{n}}=\frac{1}{2}\left(n+\frac{1}{2}\right)+\frac{1}{4}\sqrt{1+16r}\,.$$
Here the expression of $r$ is given in Eq.~(\ref{r}).  The high momentum limit of the Wigner function is discussed in detail in Appendix~\ref{app2}. Suppose ${\bf
  A}={\bf p}^m$ ($m$ positive integer) then the highest power of $p$ in the phase space
integrand for $\langle {\bf A}\rangle$ is $P=m+2(n-2j)-2$. The
momentum integral yielding $\langle {\bf A}\rangle$ will converge only
when $P<0$ yielding
\begin{eqnarray}
\sqrt{1+16 r} > m-3\,,
\label{pb}
\end{eqnarray}
giving a bound on $\beta$ when $\eta$ is a constant. When $\beta \to 0$ we have $r\to \infty$ and the above inequality holds for all values of $m$. On the other hand when $\beta$ is finite we see that $\beta$ is bounded from above for $m > 4$.

On the other hand if ${\bf A}={\bf x}^m$, where $m$ is a positive integer and ${\bf x}=i\hbar (1+\beta p^2)\partial_p$, the highest power of $p$ in the Weyl transform of ${\bf A}$ will be
$2m$ as we have seen in the Weyl transforms of ${\bf x}$ and ${\bf x}^2$. As a result, the maximum power of momenta $p$ in the integrand of the phase space integral yielding $\langle {\bf A}\rangle$  is  $P=2m+2(n-2j)-2$. The phase space integral will converge when $P<0$ which gives again an inequality
\begin{eqnarray}
\sqrt{1+16 r} > 2m-3\,,
\label{xb}
\end{eqnarray}
setting an upper limit on $\beta$. In this case also we see that if
$\beta \to 0$, the above inequality is always satisfied, irrespective
of the value of $m$. On the other hand a finite $\beta$ is restricted
as soon as $m>2$. As the harmonic oscillator Hamiltonian contains
${\bf x}^2$ we see that $\langle {\bf H}^2 \rangle$ will be calculable
in the phase space if the value of $\beta$ is below a certain
value. The integrand of the phase space integral yielding $\langle {\bf A}\rangle$  does not attenuate in the high momentum limit unless and until one restricts $\beta$ and as a result the phase space average only exists if
$\beta$ has a specific limit dependent on $m$. 

In QM following a GUP, there can be other subtleties. For pure states we generally have
$\langle {\bf A} \rangle=\bar{\bf A}$ where
\begin{eqnarray}
\bar{\bf A}  = \int \frac{\phi_n^*(p)({\bf A}\phi_n(p))}{1+\beta p^2} dp\,,
\label{aexp}
\end{eqnarray}
is the momentum space average of ${\bf A}$ and $\langle {\bf A}
\rangle$ is the phase space average of the operator as defined in
Eq.~(\ref{expa}). Here $\phi(p)$ is the momentum space wave function
of the harmonic oscillator in deformed Heisenberg algebra. When ${\bf
  A}={\bf p}^m$ for some positive integer $m$, one can easily show
that $\bar{\bf A}$ will be finite if there is an upper bound
on $\beta$ if $m>4$ and in general $\langle {\bf A} \rangle=\bar{\bf A}$. On the
other hand if ${\bf A}={\bf x}^m$ then  $\bar{\bf A}$
exists when $\beta$ has an upper bound for $m>4$. This implies $\langle {\bf A} \rangle \ne \bar{\bf A}$ for $m=3$ and 4 for arbitrary values of $\beta$ which does not satisfy the condition given in Eq.~(\ref{xb}). This discussion in general shows how the averages are dependent on the value of the deformation parameter $\beta$. For appropriate choices of $\beta$ one can always have $\langle {\bf A} \rangle=\bar{\bf A}$ for any particular $m$ and one must work with those choice of $\beta$ such that the ambiguity about the averages do not remain. 

Before we conclude this section we want to present a brief note on the experimental bounds on the deformation parameter $\beta$. There are some recent papers\cite{Bawaj:2014cda,Bushev:2019zvw} which provide experimental bounds on $\beta$ for quantum oscillators. For the harmonic oscillator we know from Eq.~(\ref{xb}) that $\beta < {\eta}/{\sqrt{6}}$ if we want to use the Wigner function, in QM modified by GUP, to find out the phase space average of ${\bf x}^4$ operator. This operator is found in the expression of the square of the Hamiltonian operator for the harmonic oscillator. This bound is dependent on two parameters, mass and frequency of the oscillator as $\eta=1/(m\hbar\omega)$. Comparing this bound to that given in Ref.~\cite{Bawaj:2014cda}, for mass of the order of $10^{-5}$\,kg, and frequency of the order of $10^3$\,Hz, the order of the bound, as obtained from our calculation, is $\beta < 10^{35}\,{\rm s}^2{\rm Kg}^{-2}{\rm m}^{-2}$. This bound yields $\beta_1 < 10^{38}$, where $\beta_1$ is a dimensionless number given in the reference, defined as $\beta_1({L_p}/{\hbar})^2=\beta$. Here $L_p=\sqrt{\hbar G/c^3}=1.6\times 10^{-35}\,{\rm m}$ is the Planck Length. The experimental upper bound, on $\beta_1$, given in Ref.~\cite{Bawaj:2014cda} for the given mass and frequency values is of the order of $10^{11}$. This shows that the experimental upper bound on $\beta$ is much less than the theoretical upper bound on it. The above discussion shows one can safely use Wigner function, as defined in our work, to calculate phase space averages of ${\bf x}^4$. Here one must also have to remember that our analysis only bounds $\beta$ when we want to use the Wigner function to get the phase space average of ${\bf x}^4$. The theoretical bound depends on the power of the operator. From Eq.~(\ref{xb}) it can be verified that if the power of the ${\bf x}$ operator increases from four the upper bound on $\beta$, obtained from phase space analysis, reduces. The low value of the experimental upper bound shows that  we can safely use the phase space averaging technique to find out the average of higher powers of position operator without contradicting any fundamental physical limits.

Comparing our bound $\beta < {\eta}/{\sqrt{6}}$ with that in Ref.~\cite{Bushev:2019zvw}, we again see that for a sapphire dumbbell oscillator with mass $m=0.3$\,kg and frequency $\omega/(2\pi) = 127$\,kHz, the theoretical bound is of the order of $\beta < 10^{27}\,{\rm s}^2{\rm kg}^{-2}{\rm m}^{-2}$. This gives $\beta_2 < 10^{29}$, where $\beta_2$ is another dimensionless number given in
Ref.~\cite{Bushev:2019zvw}, defined as $\beta_2 = \beta(M_p c)^2$. Here $M_p=\sqrt{\hbar c/G}\sim 2.2\times 10^{-8}$\,kg, is the Planck mass. The experimental upper bound on $\beta_2$ given in Ref.~\cite{Bushev:2019zvw}, for the specific choice of mass and frequency, is of the order of $10^{11}$. Using frequency measurements a more stringent bound of the order of $\beta_2<10^6$ is obtained in Ref.~\cite{Bushev:2019zvw}. From both these comparisons we see that the theoretical upper bounds on $\beta$ that we obtain is quite large in comparison to the ones obtained from experimental techniques. Here one must note that the experimental upper bounds on $\beta$ are fundamental bounds coming from actual experiments whereas the theoretical bounds on $\beta$, as predicted in the paper, arise due to application of averaging procedure in phase space. The phase space bounds on $\beta$ do not contradict the experimental bounds. The theoretical upper bound on $\beta$ depends upon the power of the position operator, and if the power of the position operator increases from four the theoretical upper bound decreases.

The above discussion shows that in general the existence of average of some dynamical operators in deformed Heisenberg algebra forces $\beta$ to have an upper bound. If $\beta \to 0$ the upper bounds disappear and we get back standard harmonic oscillator quantum mechanics where the average of all quantum operators are well defined. Before we conclude we will like to briefly opine on the elementary case of the free particle and see how the GUP affects its properties. It will become clear that the restriction of $\beta$, as shown for the harmonic oscillator, is not universal. In the case of the free particle the averages of all the operators exist for arbitrary $\beta$ values. 
%%%%%%%%%%%%%%%%%%%%%%%%%%%%%%%%%%%%%%%%%%%%%%%%%%%%%%%%%%%%%%%%%%%%%%%%%%%%%%%%%%%%%%%%%%%%%%
\section{Free Particle}
\label{fp}
%%%%%%%%%%%%%%%%%%%%%%%%

In this section we touch upon the topic of the free particle to show how a GUP actually affects the properties of the most elementary object of QM. The Hamiltonian of a free particle is
\begin{equation}
    {\bf H} = \frac{{\bf p}^2}{2m}\,.
\end{equation}
As we are doing our analysis in momentum space, the free particle wave function in deformed Heisenberg algebra is given by
\begin{equation}
    \phi(p) \propto \delta(p-p_0)\,,
\end{equation}
which is arbitrary up to a normalization constant. Using the Wigner function in the
deformed Heisenberg algebra given in Eq.~(\ref{wfgup1}) and inserting
the wave function for the free particle, we have
\begin{equation}
W = \frac{|c|^2}{h}\int\frac{du e^{\frac{ixu}{\hbar}}\delta(p-\frac{u}{2}-p_0)\delta(p+\frac{u}{2}-p_0)}{\left[1+\beta(p-\frac{u}{2})^2\right]^{\frac{1}{2}}\left[1+\beta(p+\frac{u}{2})^2\right]^{\frac{1}{2}}}\,,
\end{equation}
where $c$ is the normalization constant. Integrating the above expression we obtain
\begin{equation}
W= \frac{2|c|^2 e^{\frac{2ix(p-p_0)}{\hbar}}\delta(2(p-p_0))}{h(1+\beta p_0^2)^{1/2}[1+\beta(2p-p_0)^2]^{1/2}}\,,
\label{wfree}
\end{equation}
as the Wigner function for a free particle. Using the properties of
the delta function we can get rid of the complex exponential, but we
keep it for a later use. Although there is a complex exponential in
the Wigner function one can easily see that it is real. The Weyl
transform of the Hamiltonian is given by
\begin{equation}
     \tilde{H} = \frac{\tilde{p}^2}{2m} = \frac{p^2}{2m}\,.
\end{equation}
The phase space average of the Hamiltonian in this case is simply
\begin{equation}
    \langle {\bf H}\rangle = \int\int W(x,p)\frac{p^2}{2m} dxdp\,.
\end{equation}
After substituting the value for the Wigner function and integrating, we obtain
\begin{equation}
    \langle {\bf 
    H}\rangle=\frac{|c|^2}{2m(1+\beta p_0^2)}p_0^2\delta(0)\,,
\end{equation}
and ultimately using the value of the normalization constant we have
\begin{equation}
    \langle {\bf H}\rangle=\frac{p_0^2}{2m}\,,
\end{equation}
as expected. In the present case one can easily show that $\langle {\bf p}^m \rangle$, for positive integer $m$, exists for all $m$ without restricting $\beta$. For the position operator we have
$$\langle {\bf x} \rangle =\int \int W(x,p) x(1+\beta p^2) dx dp\,,$$
integrating the momentum variable we get
\begin{eqnarray}
\langle {\bf x} \rangle =\frac{|c|^2}{h}\int xdx=0\,.
\label{xf}
\end{eqnarray}
One can similarly show that $\langle {\bf x}^2 \rangle$ tends to infinity and as a result for the free particle with a specific momentum $p_0$, the uncertainty in position, $\Delta x$, diverges as is expected from the GUP we have been using in this paper. For the free particle the averages of ${\bf p}^m$, ${\bf x}^m$ and ${\bf H}^m$, for positive integer $m$, do not restrict the value of $\beta$. For the free particle in QM following a GUP we can easily extend the phase space picture without bothering about high momentum limit of the quasi-distribution function as in this case the Wigner function contains a Dirac delta function of momentum. The two specific cases presented in this paper shows that one can indeed define the Wigner function for QM following GUP and work with it if he/she is careful about the high momentum limit of it. If the Wigner function does not properly attenuate in the high momentum limit then one has to be careful about the value of the deformation parameter $\beta$.
%%%%%%%%%%%%%%%%%%%%%%%%%%%%%%%%%%%%%%%%%%%%%%%%%%%%%%%%%%%%%%%%%%%%%%%%%%%
\section{Discussion and conclusion}

In this paper we have tried to formulate the phase space picture of QM
which follows a particular form of a GUP. The GUP is such that the
position uncertainty is dependent on momentum uncertainty and
expectation value of the momentum in a nontrivial way. The specific
GUP arises from deforming the basic commutation relation in QM. The
new commutation relation yields a momentum dependent commutation
relation between position and momentum operator. An immediate effect of
deformed Heisenberg algebra is the absence of position eigenstates in
QM following the GUP. The GUP yields a minimal (non-zero) position uncertainty and
consequently it is impossible to measure position without a spread
forbidding position eigenstates. Momentum eigenstates are allowed in
the present case and all of our quantum mechanical analysis is
dependent on momentum space description. 

In this paper we have proposed a form of the Weyl transform of an
operator where the basic commutation relation is deformed in a particular way. We show that the proposed Weyl transform in the
deformed algebra does retain some of the most important properties of
Weyl transform of standard QM following Heisenberg uncertainty
relation. In doing so we have generalized the definition of the trace
of an operator. Once the Weyl transform has been defined we proceed to
define the Wigner function in QM following GUP. In this paper we have primarily focused on pure states as in this case, the formal
definitions can be generalized in the simplest way.  It has been shown that one can use the Wigner function and the Weyl transform of an operator, $\tilde{A}(x,p)$, and define the phase space average of the operator, $\langle {\bf A} \rangle$ for the aforementioned GUP. This is one of the most important results derived in this paper and to our understanding this is derived for the first time.

It has been shown that the Wigner function in QM following a specific form of GUP does have many of its features similar to Wigner function in QM following Heisenberg algebra. We have shown that the Wigner function in QM following GUP differs from Wigner function in standard QM in one important way. The Wigner function in the present case does not have well defined translation properties. The non existence of position basis makes the study of translation property of the Wigner function in position impossible. Moreover the Wigner function lacks translation property in the position basis also. Except this point, the expression of the Wigner function defined in the present paper is well behaved and can be taken to be the generalization of the standard Wigner function in QM.

Although the formulation of QM we have followed does not have position eigenstates because of the GUP, we can perfectly find the Weyl transform of the position operator in the present case. We have evaluated the Weyl transform of ${\bf x}$ and ${\bf x}^2$ operator using our formalism and our calculation shows that in the present case, the Weyl transform of ${\bf x}^2$ operator is not exactly the square of the Weyl transform of  ${\bf x}$. The Weyl transform of both the operators are momentum dependent and reduces to the standard QM result when the deformation parameter $\beta \to 0$. 

Then in the next section, the techniques developed in the paper are
applied to a concrete case, the case of quantum harmonic oscillator
following deformed Heisenberg algebra. The momentum space energy
eigenfunctions are known for this case from a previous work
\cite{kempf}. In the present paper we show at first that we get back
the energy eigenvalues as phase space averages using our
techniques. The slight difference between the eigenvalues and phase
space averages arise from numerical methods used to calculate the
integrals.  Another important feature of the phase space analysis of
the harmonic oscillator reveals itself by limiting the values of the
deformation parameter $\beta$ when we try to compute the phase space
averages of the powers of position and momentum operators. This is an
interesting feature in QM following GUP.  The high momentum limit of
the Wigner function may not attenuate rapidly enough for arbitrary
values of the deformation parameter $\beta$ and as a result some of
the phase space averages of operators may diverge. This divergence can
be cured if one sets an upper limit of the deformation parameter,
where the upper limit depends upon the operator whose average is being
calculated. This feature introduces subtleties in the phase space
calculation of averages of powers of operators and must be interpreted
as a limitation of QM following GUP. If the Wigner function naturally
rapidly becomes zero for high momentum limits then one may not require
to fine tune $\beta$. In the free particle case, one can calculate all
the averages without fine tuning $\beta$. To show the contrast between
physics of the harmonic oscillator and the free particle in QM
following GUP, we have briefly touched upon the topic of the free
particle in the last section.

To conclude we remind the reader that our research has shown there can
be a phase space approach to some class of QM theories following GUP
and the phase space approach works well for calculations. Although
quantum mechanically we have only momentum space wave functions, still
we can build a phase space picture of the system which is dependent on
both $x$ and $p$. In the future we hope to generalize our results to
mixed states. The nature of investigation is new and we have not seen
any other literature regarding these issues before. We hope our
research will open new avenues for quantum mechanical research into
theories following GUPs as the Wigner function approach to QM is a
very important tool for the theorists.
%%%%%%%%%%%%%%%%%%%%%%%%%%%%%%%%%%%%%%%%%%%%%%%%%%%%%%%%%%%%%%%%%%%%
\appendix\section*{\hfil Appendix \hfil}
\section{Uniqueness of Weyl transform and Wigner function}
\label{app1}
%%%%%%%%%%%%%%%%%%%%%%%%%%%%%%%%%%%%%%%%%%%%%%%
Our proof of the uniqueness of the definition of Weyl transform in QM following GUP is based on the assumption that the Weyl transform of the identity operator is unity. Keeping this in mind we now introduce a general function $f(x,p,u,\beta)$ and define the Weyl transform in deformed Heisenberg Algebra as :
\begin{equation}
\tilde{A}(x,p) = \int du e^{\frac{ixu}{\hbar}}f(x,p,u,\beta)\langle p+u/2|{\bf A}|p-u/2\rangle\,.
\end{equation} 
Our assumption $\tilde{I} = 1$ demands 
\begin{equation}
    \tilde{I} = \int du e^{\frac{ixu}{\hbar}}f(x,p,u,\beta)\langle p+u/2|{\bf I}|p-u/2\rangle = 1\,.
\end{equation}
The above integration can easily be done showing that 
\begin{equation}
f(p,u=0,\beta) = \frac{1}{(1+\beta p^2)}\,.
\end{equation}
The above form of $f$ does not depend upon $x$. Henceforth we assume that in general $f$ only depends on $p$ and $u$. Now, we want our Weyl transform to satisfy the property
\begin{equation}
\frac{1}{h}\int\int\tilde{A}\tilde{B}dxdp= {\rm Tr}[{\bf A} {\bf B}]\,.
\label{appn1}  
\end{equation}
This is very important for our analysis based on the momentum states as it is the key equation to calculate the averages of observable. The left hand side of the above equation can be written as:
\begin{eqnarray}
  \frac{1}{h}\int\int\tilde{A}\tilde{B}dxdp &=& \frac{1}{h}\int \int \int \int e^{i\frac{(u_1 + u_2)x}{\hbar}}\langle p+ u_1/{2|{\bf A}|p- u_1/2\rangle\langle p+ u_2/2|{\bf B}|p- u_2/2\rangle}\nonumber\\
&\times& f(\beta, p, u_1)f(\beta, p, u_2)\, du_1 \,du_2 \,dx \,dp\,.
\nonumber  
\end{eqnarray}
Integrating with respect to $x$ we get a delta function 
\begin{equation}
    \delta(u_1 + u_2) = \frac{1}{2\pi \hbar}\int e^{\frac{i(u_1 + u_2)x}{\hbar}} dx\,,
\end{equation}
using which we can now write,
\begin{eqnarray}
  \frac{1}{h}\int\int\tilde{A}\tilde{B}dxdp &=&\int \int \int \delta(u_1+u_2)\langle p+ u_1/2|{\bf A}|p- u_1/2\rangle\langle p+ u_2/2|{\bf B}|p- u_2/2\rangle \nonumber\\
&\times& f(\beta, p, u_1)f(\beta, p, u_2) \,dp \,du_1\,du_2\nonumber\\
&=& \int\int \langle p+ u_1/2|{\bf A}|p- u_1/2\rangle \langle p- u_1/2 |{\bf B}|p+ u_1/2 \rangle f(\beta, p, u_1)f(\beta, p, -u_1) du_1 dp\,.
\nonumber  
\end{eqnarray}
Defining new variables as $p + \frac{u_{1}}{2} = v $ and $p - \frac{u_{1}}{2} = y$
such that $dp du_1 = dvdy$ the above expression transforms to
\begin{equation}
  \frac{1}{h}\int\int\tilde{A}\tilde{B}dxdp= \int\int\langle v|{\bf A}|y\rangle\langle y|{\bf B}|v\rangle f(\beta, p, 2(v-y))f(\beta, p, -2(v-y)) dy\,dv\,.
\label{appn2}  
\end{equation}
Now, the right hand side of Eq.~(\ref{appn1}) is 
\begin{equation}
    {\rm Tr}[{\bf A \bf B}] \equiv \int \frac{\langle v|{\bf A \bf B}|v \rangle}{1+\beta v^2} dv=\int\frac{\langle v|{\bf A}|y\rangle\langle y|{\bf B}|v\rangle} {(1+\beta v^2)(1+\beta y^2)}\,dv\,dy\,.
\end{equation}
We know that if Eq.~(\ref{appn2}) is equal to the right hand side of Eq.~(\ref{appn1}), then from the last equation above we must have
\begin{equation}
    f(\beta, p, 2(v-y))f(\beta, p, -2(v-y))=\frac{1}{(1+\beta v^2)(1+\beta y^2)}\,.
\end{equation}
The above equation yields
\begin{equation}
    f(\beta,p,u_1)f(\beta,p,-u_1) = \frac{1}{\left[1+\beta(p+u_1/2)^2\right]\left[1+\beta(p-u_1/2)^2\right]}\,.
\end{equation}
This equation provides three different possible forms of the function $f(\beta,p,u)$ as: 
\begin{equation}
    f(p,u,\beta) = \frac{1}{\left[1+\beta(p+u/2)^2\right]}\,,
\end{equation}
or 
\begin{equation}
    f(p,u,\beta) = \frac{1}{\left[1+\beta(p-u/2)^2\right]}\,,
\end{equation}
or
\begin{equation}
    f(p,u,\beta) = \frac{1}{\left[1+\beta(p+u/2)^2\right]^{\frac{1}{2}}\left[1+\beta(p-u/2)^2\right]^{\frac{1}{2}}}\,.
\end{equation}
The Weyl transform of {\bf x} ($\tilde{x}$) using the above forms of $f(p,u,\beta)$ are given respectively by: 
\begin{equation}
    \tilde{x} = x(1+\beta p^2) - i\hbar\beta p\,,
\end{equation}
\begin{equation}
    \tilde{x} = x(1+\beta p^2) + i\hbar\beta p\,,
\end{equation}
and
\begin{equation}
    \tilde{x} = x(1+\beta p^2)\,.
\end{equation}
But, Weyl transform of observable (Hermitian operators) is purely real, so we can disregard the form of $f(p,u,\beta)$ which produces complex Weyl transforms. Thus we get the unique form of the Weyl transform to be: 
\begin{equation}
    \tilde{A}(x,p) = \int du e^{\frac{ixu}{\hbar}}\frac{\langle p+u/2|{\bf A}|p-u/2\rangle}{\left[1+\beta(p+u/2)^2\right]^{\frac{1}{2}}\left[1+\beta(p-u/2)^2\right]^{\frac{1}{2}}}\,. 
\end{equation} 
This is exactly the Weyl transform we have used in the paper.
%%%%%%%%%%%%%%%%%%%%%%%%%%%%%%%%%%%%%%%%%%%%%%%%%%%%%%%%%%%%%%%%%%%%%%%%%%%%%%%%%%%%%%%%%%%
\section{High momentum behavior of the Wigner function}
\label{app2}
%%%%%%%%%%%%%%%%%%%%%%%%%%%%%%%%%%%%%%%%%%%%%%%%%%%%%%%%%%%%%%%%%%%%%%%%%%%%%%%%%%%%%%%%%%%
Wigner function is given by : 
\begin{equation}
W(x,p) = \frac{1}{h}\int e^{\frac{ix u}{\hbar}} \frac{\phi(p+ u/2)\phi ^*(p- u/2)}{\left[1+\beta (p-\frac{u}{2})^{2}\right]^{\frac{1}{2}}\left[1+\beta(p+\frac{u}{2})^{2}\right]^{\frac{1}{2}}}du\,.
%\label{wfgup1}
\end{equation}
Let power of $p$ in the wave function $\phi$ be $m$. Thus for large $p$, the Wigner function $W(x,p)$ will go as $p^{2m-2}$. Now, for harmonic oscillator, $\phi$ takes the form : 
\begin{equation}
\phi_{n}(p) \propto \frac{1}{(1+\beta p^{2})^{\sqrt{q+r_{n}}}}\,F\left(a_{n},-n,c_{n};\frac{1+i\sqrt{\beta}p}{2}\right)\,,
%\label{wfho}  
\end{equation}
where $F\left(a_{n},-n,c_{n};\frac{1+i\sqrt{\beta}p}{2}\right)$ is the Hypergeometric polynomial function or order $n$. From these facts we can write $m = n-2\sqrt{q+r_n} = n-2j$ where $j \equiv
\sqrt{q+r_{n}}=\frac{1}{2}\left(n+\frac{1}{2}\right)+\frac{1}{4}\sqrt{1+16r}$. Therefore $W(x,p) \propto p^{2(n-2j)-2}$ in the high momentum limit.
%%%%%%%%%%%%%%%%%%%%%%%%%%%%%%%%%%%%%%%%%%%%%%%%%%%%%%%%%%%%%%


\begin{thebibliography}{15}

\bibitem{case} 
William B. Case,
\textit{Wigner functions and Weyl transforms for pedestrians}. 
Am. J. Phys. 76(10)(2008)

\bibitem{kempf} 
Achim Kempf, Gianpiero Mangano, Robert B. Mann,
\textit{Hilbert Space Representation of the Minimal Length Uncertainty Relation}. 
Phys.Rev. D52:1108 (1995)

\iffalse
\bibitem{case} 
William B. Case,
\textit{Wigner functions and Weyl transforms for pedestrians}. 
Am. J. Phys. 76(10)(2008)
\fi

\bibitem{maggore} 
 M. Maggiore,
\textit{Quantum groups, gravity, and the generalized uncertainty principle}.
Phys.Rev. D49, 5182 (1994).

\bibitem{kempf2} 
Achim Kempf,
\textit{Uncertainty relation in quantum mechanics with quantum group symmetry}. 
J. Math. Phys. 35, 4483 (1994)

\bibitem{kempf3} 
Achim Kempf, \textit{Quantum group-symmetric Fock spaces with Bargmann-Fock representation}. Lett. Math. Phys. 26, 1 (1992).

\bibitem{kempf4} 
Achim Kempf,
\textit{Quantum group symmetric Bargmann–Fock space: Integral kernels, Green functions, driving forces}. J. Math. Phys.34, 969 (1993).


\bibitem{kempf5} 
Achim Kempf,
\textit{Quantum Field Theory with Nonzero Minimal Uncertainties in Positions and Momenta}. Preprint DAMTP/94-33, hep-th/9405067 (1994).


\bibitem{brau} 
Fabian Brau, Fabian Buisseret, 
\textit{Minimal length uncertainty relation and gravity quantum well}. 
Phys.Rev. D52:1108 (2006)



\bibitem{jaeckel} 
M-T. Jaeckel, S. Reynaud,
\textit{Gravitational quantum limit for length measurements}. 
Phys. Lett. A185, 143 (1994)

\bibitem{lizzi} 
 A.P. Balachandran, G. Bimonte, E. Ercolessi, G. Landi, F. Lizzi, G. Sparano, P. Teotonio-Sobrinho,
\textit{Finite quantum physics and non commutative geometry}. 
Preprint ICTP: IC/94/38, DSF-T-2/94, SU-4240-567 (1994).

\bibitem{maggiore} 
 M. Maggiore,
\textit{The algebraic structure of the generalized uncertainty principle}. 
Phys. Lett. B319, 83 (1993).

\bibitem{konishi} 
 K. Konishi, G. Paffuti, P. Provero,
\textit{Minimum physical length and the generalized uncertainty principle in string theory}. 
Phys. Lett. B234, 276 (1990).


\bibitem{amati} 
 D. Amati, M. Cialfaloni, G. Veneziano,
\textit{Can space time be probed below the string size}. 
Phys. Lett. B216, 41 (1989).

\bibitem{wigner} 
 E. Wigner,
\textit{On the quantum correction for thermodynamic equilibrium}.
Phys. Rev. 40, 749–759 (1932).

\bibitem{hillery} 
M. Hillery, R. F. O’Connell, M. O. Scully, and E. P. Wigner,
\textit{Distribution functions in physics: Fundamentals}.
Phys. Rep. 106, 121–167 (1984).

\bibitem{lee} 
 H.Lee,
\textit{Theory and application of the quantum phase-space distribution functions}.
Phys. Rep. 259, 150–211 (1995).

\bibitem{weyl} 
 H.Weyl,
\textit{The Theory of Groups and Quantum Mechanics} (Dover, New York, 1931)

%\cite{Bosso:2018uus}
\bibitem{Bosso:2018uus}
P.~Bosso,
%``Rigorous Hamiltonian and Lagrangian analysis of classical and quantum theories with minimal length,''
Phys. Rev. D \textbf{97}, no.12, 126010 (2018)
doi:10.1103/PhysRevD.97.126010
[arXiv:1804.08202 [gr-qc]].
%10 citations counted in INSPIRE as of 06 Nov 2020

%\cite{Bosso:2020aqm}
\bibitem{Bosso:2020aqm}
P.~Bosso,
%``On the quasi-position representation in theories with a minimal length,''
[arXiv:2005.12258 [gr-qc]].
%0 citations counted in INSPIRE as of 06 Nov 2020

%\cite{Das:2014bba}
\bibitem{Das:2014bba}
S.~Das, M.~P.~G.~Robbins and M.~A.~Walton,
%``Generalized Uncertainty Principle Corrections to the Simple Harmonic Oscillator in Phase Space,''
Can. J. Phys. \textbf{94}, no.1, 139-146 (2016)
doi:10.1139/cjp-2015-0456
[arXiv:1412.6467 [gr-qc]].
%12 citations counted in INSPIRE as of 06 Nov 2020

%\cite{Bosso:2017ndq}
\bibitem{Bosso:2017ndq}
P.~Bosso, S.~Das and R.~B.~Mann,
%``Planck scale Corrections to the Harmonic Oscillator, Coherent and Squeezed States,''
Phys. Rev. D \textbf{96}, no.6, 066008 (2017)
doi:10.1103/PhysRevD.96.066008
[arXiv:1704.08198 [gr-qc]].
%15 citations counted in INSPIRE as of 07 Nov 2020

%\cite{Bosso:2018syo}
\bibitem{Bosso:2018syo}
P.~Bosso and S.~Das,
%``Generalized ladder operators for the perturbed harmonic oscillator,''
Annals Phys. \textbf{396}, 254-265 (2018)
doi:10.1016/j.aop.2018.07.022
[arXiv:1807.05436 [quant-ph]].
%4 citations counted in INSPIRE as of 07 Nov 2020

%\cite{Bawaj:2014cda}
\bibitem{Bawaj:2014cda}
M.~Bawaj, C.~Biancofiore, M.~Bonaldi, F.~Bonfigli, A.~Borrielli, G.~Di Giuseppe, L.~Marconi, F.~Marino, R.~Natali and A.~Pontin, \textit{et al.}
%``Probing deformed commutators with macroscopic harmonic oscillators,''
Nature Commun. \textbf{6}, 7503 (2015)
doi:10.1038/ncomms8503
[arXiv:1411.6410 [gr-qc]].
%51 citations counted in INSPIRE as of 03 Jan 2021

%\cite{Bushev:2019zvw}
\bibitem{Bushev:2019zvw}
P.~A.~Bushev, J.~Bourhill, M.~Goryachev, N.~Kukharchyk, E.~Ivanov, S.~Galliou, M.~E.~Tobar and S.~Danilishin,
%``Testing the generalized uncertainty principle with macroscopic mechanical oscillators and pendulums,''
Phys. Rev. D \textbf{100}, no.6, 066020 (2019)
doi:10.1103/PhysRevD.100.066020
[arXiv:1903.03346 [quant-ph]].
%19 citations counted in INSPIRE as of 03 Jan 2021

\end{thebibliography}
\end{document}